\documentclass[sigconf,natbib=true]{acmart}

\usepackage{color}

\newcommand{\heading}[1]{\vspace*{1mm}\noindent\textbf{#1.}}
\AtBeginDocument{%
  \providecommand\BibTeX{{%
    \normalfont B\kern-0.5em{\scshape i\kern-0.25em b}\kern-0.8em\TeX}}}

\usepackage{bbm}
\usepackage{multirow}
\usepackage[inline]{enumitem}
\usepackage{graphicx}
\usepackage{sistyle}
\usepackage[ruled]{algorithm2e}
\usepackage{booktabs}
\usepackage{caption}
\SIthousandsep{,}
\usepackage{makecell}
\usepackage[skip=0pt]{caption}
\usepackage{amsmath}
\usepackage{subfigure}
\usepackage{hyperref}
\usepackage{array} 
\usepackage{graphicx}
\usepackage{diagbox}
\usepackage{textcomp}
\usepackage[normalem]{ulem}

\makeatletter
\g@addto@macro\normalsize{%
  \abovedisplayskip 0pt plus1pt 
  \belowdisplayskip 0pt plus1pt
  \abovedisplayshortskip  0pt plus1pt%
  \belowdisplayshortskip  0pt plus1pt
}
\makeatother

\copyrightyear{2023}
\acmYear{2023}
\acmConference[Gen-IR@SIGIR2023]{The First Workshop on Generative Information Retrieval}{July 27, 2023}{Taipei, Taiwan}
\acmBooktitle{The First Workshop on Generative Information Retrieval (Gen-IR@SIGIR23), July 27, 2023, Taipei, Taiwan}
\acmDOI{}
\acmPrice{}
\acmISBN{}
\setcopyright{rightsretained}

\begin{document}

\title{On the Robustness of Generative Retrieval Models: An Out-of-Distribution Perspective}

\author{Yu-An Liu}
\orcid{0000-0002-9125-5097}
\author{Ruqing Zhang}
\authornote{Research conducted when the author was at the University of Amsterdam.}
\orcid{0000-0003-4294-2541}
\affiliation{
	\institution{CAS Key Lab of Network Data Science and Technology, ICT, CAS}
	\institution{University of Chinese Academy of Sciences}
	\city{Beijing}
	\country{China}
}
\email{{liuyuan21b,zhangruqing}@ict.ac.cn}

\author{Jiafeng Guo}
\orcid{0000-0002-9509-8674}
\authornote{Jiafeng Guo is the corresponding author.}
\author{Wei Chen}
\orcid{0000-0002-7438-5180}
\affiliation{
	\institution{CAS Key Lab of Network Data Science and Technology, ICT, CAS}
	\institution{University of Chinese Academy of Sciences}
	\city{Beijing}
	\country{China}
}
\email{{guojiafeng,chenwei2022}@ict.ac.cn}

\author{Xueqi Cheng}
\orcid{0000-0002-5201-8195}
\affiliation{
	\institution{CAS Key Lab of Network Data Science and Technology, ICT, CAS}
	\institution{University of Chinese Academy of Sciences}
	\city{Beijing}
	\country{China}
}
\email{cxq@ict.ac.cn}

\renewcommand{\shortauthors}{Yu-An Liu et al.}

\begin{abstract}

Recently, we have witnessed generative retrieval increasingly gaining attention in the information retrieval (IR) field, which retrieves documents by directly  generating their identifiers. 
So far, much effort has been devoted to developing effective generative retrieval models. 
There has been less attention paid to the robustness perspective. 
When a new retrieval paradigm enters into the real-world application, it is also critical to measure the out-of-distribution (OOD) generalization, i.e., how would generative retrieval models generalize to
new distributions. 
To answer this question, firstly, we define OOD robustness from three perspectives in retrieval problems: 1) The \textit{query variations}; 2) The \textit{unforeseen query types}; and 3) The \textit{unforeseen tasks}. 
Based on this taxonomy, we conduct empirical studies to analyze the OOD robustness of several representative generative retrieval models against dense retrieval models.
The empirical results indicate that the OOD robustness of generative retrieval models requires enhancement.
We hope studying the OOD robustness of generative retrieval models would be advantageous to the IR community.

\end{abstract}



\maketitle

\section{Introduction}
Large-scale document retrieval has gained widespread attention due to its benefits to many real-world applications \cite{mitra2017learning, chen2017survey,liu2023topicoriented}. 
With the development of representation learning techniques \cite{devlin2018bert}, major progress has turned to dense retrieval based on the ``index-retrieve'' pipeline, which suffers from a large memory footprint and difficulty in end-to-end optimization. 
Recently, \citet{metzler2021rethinking} envisioned a fundamentally generative retrieval paradigm, to fully parameterize different components of index and retrieval with a single consolidated model. 
Specifically, a sequence-to-sequence (seq2seq) learning framework is employed to directly predict the identifiers of relevant documents (docids) with respect to a given query. 

Current research on generative retrieval is often studied in homogeneous and narrow settings, that is, assuming the train and test examples are independent and identically distributed (I.I.D.). 
Under the I.I.D. assumption, proposed generative retrieval models have achieved promising performance on large-scale document retrieval tasks \cite{chen2022corpusbrain,tay2022transformer, wang2022neural}. 
However, in real scenarios, the I.I.D. assumption can hardly be satisfied, where the testing distribution is usually unknown and different from the training. 
High I.I.D. accuracy does not necessarily translate to out-of-distribution (OOD) robustness for document retrieval. 
Besides, pre-trained Transformers, which usually serve as the backbone of existing generative retrieval models, may rely on spurious cues and annotation artifacts are less likely to include OOD examples \cite{hendrycks2020pretrained}.
In this way, the OOD robustness of generative retrieval models remains uncertain. 

In this work, we systematically study the OOD robustness of several representative generative retrieval models. 
We decompose OOD robustness into a model's generalization ability to (1) Query variations, (2) Unforeseen query types, and  (3) Unforeseen tasks.
Each generalization ability perspective corresponds to a different OOD scenario in reality.
Based on this taxonomy, we design corresponding experiments and conduct empirical studies to analyze the robustness of several representative generative retrieval models against dense retrieval models.

For the experiment, we employ the comprehensive knowledge-intensive language tasks (KILT) benchmark \cite{petroni-etal-2021-kilt}, comprising eleven datasets across five distinct KILT tasks. 
Along with multiple corpora under each task, it is ideal for the initial analysis of OOD robustness. 
In the future we will also try to explore more general datasets and models for experimentation.
In this work, following \cite{chen2022corpusbrain, deautoregressive}, we consider the retrieval task of KILT, in which the model should retrieve a set of Wikipedia pages as evidence for the final prediction with respect to the input query. 
The results reveal generative retrieval models' overall poor performance in OOD robustness, and they have different generalizability performance in different OOD scenarios.
As a result, there is considerable scope for future robustness improvements. 
To facilitate future research in this area, we release the code and data at \url{https://github.com/ict-bigdatalab/GR_OOD}.

\section{Related Work}
\textbf{Dense retrieval models}. Dense retrieval models typically adopt a bi-encoder architecture to encode queries and documents into low-dimension embeddings
and utilize embedding similarities as estimated relevance scores for effective retrieval~\cite{guo2022semantic}. 
\citet{Karpukhin2020DensePR} were pioneers in discovering that fine-tuning BERT to learn effective dense representations, called DPR, outperforms traditional retrieval methods like BM25.
Subsequently, researchers began exploring various fine-tuning techniques to enhance dense retrieval models, such as mining hard negatives~\cite{Xiong2021ApproximateNN,Zhan2021OptimizingDR}, late interaction~\cite{Khattab2020ColBERTEA}.
Recently, researchers have also investigated pre-training tasks for dense retrieval~\cite{gao2021condenser, ma2022contrastive}.
Although these methods greatly improve the performance of dense retrieval models, they follow the same bi-encoder architecture represented by the DPR and usually need considerable memory consumption and computational overhead.

\noindent \textbf{Generative retrieval models}. Generative retrieval has recently garnered increasing interest \cite{metzler2021rethinking,chen2022gere,wang2022neural,bevilacqua2022autoregressive}, which retrieves documents by directly generating their identifiers based on the given query.
It offers an end-to-end solution for document retrieval tasks \cite{metzler2021rethinking, chen-2023-unified} and allows for better exploitation of the capabilities of large generative language models.
For example, 
\citet{deautoregressive} proposed an autoregressive entity retrieval model and  
\citet{tay2022transformer} introduced a differentiable search index (DSI) and represent documents as atomic ids, naive string, or semantic strings.
\citet{chen2022corpusbrain} proposed a pre-trained generative retrieval model called CorpusBrain to encode all information of the corpus within its parameters in a general way. 
However, the robustness of generative retrieval models has been overlooked by the community.

\noindent \textbf{Out-of-distribution in IR}. Current studies on OOD robustness in IR have their own limitations.
For example, \citet{wu2022neural} only explored the OOD generalization performance of neural ranking models.
Some works have been devoted to alleviating the poor performance of dense retrieval in the scenarios of query variants \cite{zhuang2021dealing,penha2022evaluating,chen2022towards,sidiropoulos2022analysing} or zero/few-shot of corpus \cite{yu2022coco,liang2020embedding,thakur2beir}.
In this work, we focus on the OOD generalizability of generative retrieval models.

\vspace*{-2mm}
\section{I.I.D Setting of Retrieval Problem}

For a better description of the OOD setting of the retrieval problem, we first briefly introduce the I.I.D setting of the retrieval problem. 

Formally, given a dataset $\mathcal{D} = \{(q_i, D, Y_i)\}^{n}_{i=1}$, where $q_i$ denotes a query, $D = \{d_{1}, d_{2}, ..., d_{N}\}$ represents the corpus, and $Y = \{r_1, r_2, ..., r_l\}$ indicates the corresponding relevance label of each document in $D$.
Total order exists among the relevance labels such that $r_l \succ r_{l-1} \succ ... \succ r_1$, where $\succ$ denotes the order relation. 
Each query $q_i$ is associated with a list of corresponding labels $\mathbf{y}_i = \{y_{i1}, y_{i2}, ... ,y_{i, N}\}$, where $N$ denotes the corpus size.

Traditionally, a retrieval model could be a term-based retrieval mode\cite{StephenRobertson1994SomeSE,ponte2017language} or a dense retrieval model \cite{Karpukhin2020DensePR,ma2022contrastive}.
Recently, generative retrieval models have emerged as another paradigm \cite{chen2022corpusbrain,tay2022transformer,deautoregressive}.
Although the paradigm is different, these retrieval models have the same formal definition under the retrieval task.
Without loss of generality, we use $f$ to denote the retrieval model.
We consider the retrieval model $f$ learned on the dataset $\mathcal{D}$, which is drawn from the training distribution $\mathcal{G}$. 
The goal of the retrieval is to employ the learned model $f$ to generate a score for any query-document pair $(q,d)$, reflecting the relevance degree between $q$ and $d$, and thus allows one to produce a permutation $\pi(q_t, D, f)$ according to predicted scores.
Given an effectiveness evaluation metric $M$, retrieval models are typically evaluated by the average performance over the test queries under the I.I.D. setting, i.e.,
\begin{equation}
	\mathbb{E}_{(q_t,D,\mathbf{y}_t)\sim\mathcal{G}} M(\pi(q_t, D, f), \mathbf{y}_t),
\label{equ: mean_performance}
\end{equation} 
where $q_t,D$ and $\mathbf{y}_t$ denotes the query, the corpus and the label in the test set, respectively. Specifically, the test samples are supposed to be drawn from the same distribution as $\mathcal{G}$. 

\vspace*{-2mm}
\section{OOD Setting of Retrieval Problem}
In this work, we define the OOD robustness of retrieval models in three ways, i.e., query variations, unforeseen query types and unforeseen tasks. 
For query variations, the models are trained on the original dataset $\mathcal{D}$ and tested on the same dataset with query variations.
For unforeseen query types and unforeseen tasks, the models are trained on the original dataset $\mathcal{D}$ and tested on the new dataset $\mathcal{D}'$ from the same task with $\mathcal{D}$ and $\tilde{\mathcal{D}}$ whose task is different from $\mathcal{D}$, respectively.

\heading{Query Variations}
The query variations refer to different expressions of the same information need.
Therefore, a query and its variations usually correspond to the same related document. 
This query-level OOD aims to analyze the model's generalizability across different query variations within the dataset.

Formally, suppose that the examples $q_t, D$ and $\mathbf{y}_t$ are drawn from the training distribution $\mathcal{G}$. 
We aim to evaluate the models' performance on the query OOD example. 
Specifically, the testing scenario of OOD generalizability on query variations is defined as
\begin{equation}
	\mathbb{E}_{(q_t, D,\mathbf{y}_t)\sim\mathcal{G}} M(\pi(G(q_t), D, f), \mathbf{y}_t),
\label{equ: defensive ability against query attack}
\end{equation} 
where $G(q_t)$ is the query variations generated by the generator $G$.

\heading{Unforeseen Query Types}
The unforeseen query types refer to the unseen types of queries due to new types of information needs on the same task.
Due to the query-specific provenance, query distributions differ between one query set to another, every though they focus on the same task.
This query-type-level OOD aims to analyze the model's generalizability across different query types.

Formally, suppose that the new types of queries OOD examples $q_t'$ and relevance label $\mathbf{y}_t'$ are drawn from the new distribution $\mathcal{G}'_Q$ and come from dataset $\mathcal{D}'$. 
The corpus of datasets $\mathcal{D}$ and $\mathcal{D}'$ are consistent as $D$.
Specifically, the testing scenario of OOD generalizability on unforeseen query types is defined as 
\begin{equation}
	\mathbb{E}_{(q_t',D,\mathbf{y_t'})\sim\mathcal{G}'_Q} M(\pi(q_t', D, f) \mathbf{y}_t'),
\label{equ: OOD generalizability on unforeseen type}
\end{equation}
note that the training dataset $\mathcal{D}$ and the testing dataset $\mathcal{D}'$ with unseen query types come from the same task. 

\heading{Unforeseen Tasks}
The unforeseen tasks refer to distribution shifts arising from task shifts.
In practice, a retrieval model is usually trained to focus on a specific task and model a particular relevance pattern.  
Therefore, it is essential to evaluate how well a retrieval model, trained on datasets of a given task, can generalize to datasets of new tasks.
This pair-level OOD aims to analyze the model's generalizability across different retrieval tasks.

Formally, suppose that the new task OOD examples $\tilde{q}_t$, $\tilde{D}$ and $\tilde{\mathbf{y}}_t$ are drawn from the new distribution $\tilde{\mathcal{G}_T}$ and come from dataset $\tilde{\mathcal{D}}$. 
Specifically, the testing scenario of OOD generalizability on unforeseen corpus is defined as 
\begin{equation}
	\mathbb{E}_{(\tilde{q}_t,\tilde{D},\tilde{\mathbf{y}}_t)\sim\tilde{\mathcal{G}}_T} M(\pi(\tilde{q}_t, \tilde{D}, f), \tilde{\mathbf{y}}_t),
\label{equ: OOD generalizability on unforeseen task}
\end{equation}
note that the training dataset $\mathcal{D}$ and the testing dataset $\tilde{\mathcal{D}}$ belong to different tasks, respectively. 

\vspace*{-2mm}
\section{Experimental Setup}
\subsection{Datasets}

\begin{figure}[t]
    \centering
    \includegraphics[scale=0.67]{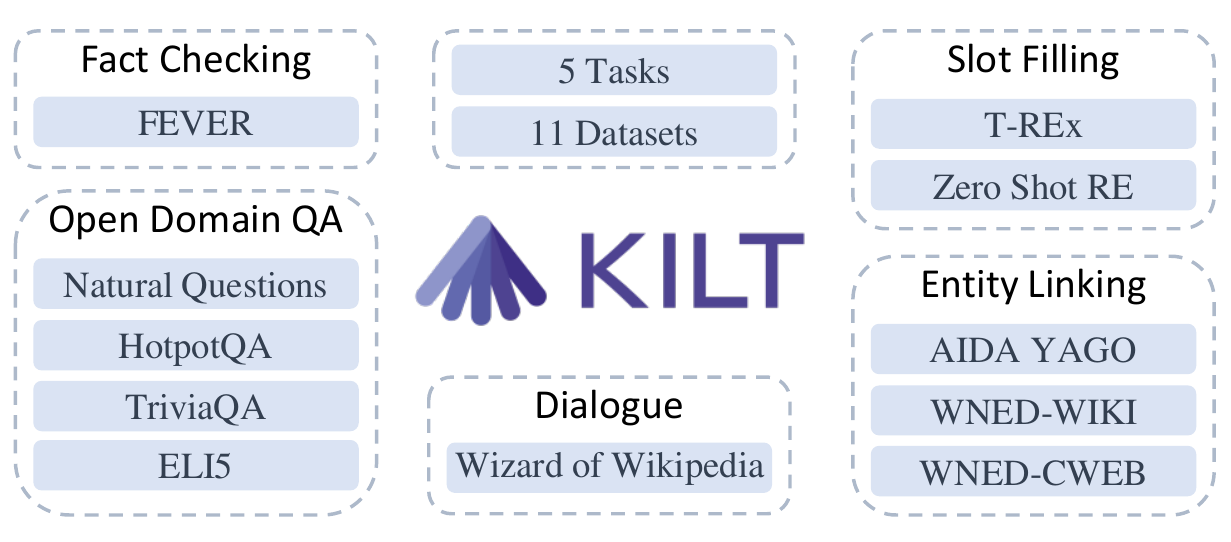}
    \caption{An overview of the diverse tasks and datasets considered in KILT benchmark.}
    \label{fig:TASK STATEMENT}
\end{figure}

For three OOD settings, we construct three benchmark datasets based on the KILT benchmark \cite{petroni-etal-2021-kilt} (see Figure \ref{fig:TASK STATEMENT}). 
Due to the submission frequency limits of the online leaderboard, we used the performance on the dev set to evaluate model performance. 
In the following, we describe the details of the constructed datasets for evaluating the OOD generalizability on query variation, unforeseen query types, and unforeseen tasks, respectively. 
\begin{itemize}[leftmargin=*]
    \item \textbf{Dataset for query variations}. We use the queries in Fever (FEV) and Natural Questions (NQ) to generate their variations, as all of the retrieval models perform relatively well on these datasets. Four generation strategies are considered \cite{penha2022evaluating} to perturb input queries, including (1) \textbf{Misspelling} for randomly substituting existing characters; (2) \textbf{Naturality} for removing all stop words; (3) \textbf{Order} for randomly exchanging positions of two words; and (4) \textbf{Paraphrasing} for replacing non-stop words according to the similarity of counter-fitted word embeddings \cite{NikolaMrki2016CounterfittingWV}.
Examples of the generated query variations are listed in Table \ref{table:query variation}.

\item \textbf{Dataset for unforeseen query types}. We utilize the datasets under the open-domain QA task which covers the largest number of datasets in the KILT.
There are three full datasets in open domain QA, i.e., NQ, HoPo, TQA. 
These datasets contain different topics and provenances, i.e., web search queries \cite{nq}, multi-hop questions \cite{hotpotqa}, and trivia questions \cite{triviaqa}.

\item \textbf{Dataset for unforeseen tasks}. We use 5 tasks from the KILT benchmark, namely, fack checking (FC), entity linking (EL), slot filling (SF), open domain question answering (QA), and dialogue (Dial.).
For each task in the KILT, we mix every training and test set of all datasets under each task separately to create a new dataset for that task.
\end{itemize}

\subsection{Retrieval models}
\begin{itemize}[nosep,leftmargin=*]
\item \textbf{DPR}~\cite{Karpukhin2020DensePR} is a representative dense retrieval model which uses dual-encoder architecture and is trained with in-batch negatives and a few hard negatives selected with BM25. 
\item \textbf{BART}~\cite{lewis2020bart} is a Seq2Seq model applicable for sequence generation tasks. 
Following \cite{deautoregressive,chen2022corpusbrain}, we extract the query-title pairs from each dataset and fine-tune the BART for generative retrieval.
\item \textbf{CorpusBrain}~\cite{chen2022corpusbrain} (C.B. for short) is a pre-trained generative retrieval model for knowledge-intensive language tasks.
We fine-tune CorpusBrain on every specific downstream KILT task.
\end{itemize}

\begin{table}[t]
\centering
 \renewcommand{\arraystretch}{1}
 \setlength\tabcolsep{1.8pt}
 \caption{Synthetic queries using variation generators.}
 \small
\begin{tabular}{cl} 
\toprule
Original query & who wrote most of the declaration of independence  \\
\hline
Misspelling & who wr\textbf{eit} most of the declaration of independence  \\
Naturality & \textbf{\sout{who}} wrote  most \textbf{\sout{of}} \textbf{\sout{the}} declaration \textbf{\sout{of}} independence \\
Order & who \textbf{declaration} most of the \textbf{wrote} of independence \\
Paraphrasing & who \textbf{authored} most of the declaration of independence \\
 \bottomrule
\end{tabular}
\label{table:query variation}
\end{table}

\begin{table}[t]
\centering
   \caption{R-precision (\%) for the page-level retrieval task on the KILT dev data.}
   \renewcommand{\arraystretch}{0.85}
   \setlength\tabcolsep{1.5pt}
  	\begin{tabular}{l  c c c c c c c c | c }
  \toprule
  \multirow{2}{*}{\textbf{Model}} & \multicolumn{1}{c}{\textbf{FC}} & \multicolumn{1}{c}{\textbf{EL}} & \multicolumn{2}{c}{\textbf{Slot Filling}} & \multicolumn{3}{c}{\textbf{Open Domain QA}} & \multicolumn{1}{c}{\textbf{Dial.}} & \\ 
       & \textbf{FEV} & \textbf{AY2} & \textbf{T-REx} & \textbf{zxRE} & \textbf{NQ} & \textbf{HoPo} & \textbf{TQA} & \textbf{WoW} & \textbf{Avg.} \\ 
       \midrule
DPR  & 52.9 & \phantom{1}2.1 & 13.5 & 28.4 & 53.9 & 26.1 & 42.8 & 25.2 & 30.6\\
BART  & 79.6 & 90.1 & 74.4 & 94.3 & 48.9 & 41.6 & 64.4 & 50.7 & 68.0\\
C.B.  & \textbf{81.4} & \textbf{90.7} & \textbf{75.7} & \textbf{97.6} & \textbf{57.6} & \textbf{50.7} & \textbf{70.9} & \textbf{55.0} & \textbf{72.5}\\
\bottomrule
    \end{tabular}
   \label{table:Baseline}
\end{table}

\vspace{-1mm}
\subsection{Evaluation}
To measure the OOD generalizability of the retrieval models, following \cite{wu2022neural}, we use \textbf{DR$_{OOD}$} (\%) to evaluate the drop rate between the retrieval performance $P_{OOD}$ under the OOD setting 
and the retrieval performance $P_{I.I.D.}$ under the I.I.D. setting, defined as,
\begin{equation} 
\label{DROOD}
	DR_{OOD} = \frac{P_{OOD} - P_{I.I.D.}}{P_{I.I.D.}},
\end{equation} 
where $P_{I.I.D.}$ denotes the retrieval performance of the model trained on the  training set corresponding to the test set.
And $P_{OOD}$ denotes the retrieval performance of the model trained on the  training set that is out-of-distribution for the test set. 
The ranking model would be more robust with a higher DR$_{OOD}$. 

The effectiveness metric for evaluating the retrieval performance in KILT is usually defined as \textbf{R-precision} (\%), which is suggested in the official instructions and widely used in previous works on KILT \cite{deautoregressive, chen2022corpusbrain, chen2022gere}. 
R-precision is calculated as $\frac{r}{R}$, where $R$ is the number of Wikipedia pages inside each provenance set and $r$ is the number of relevant pages among the top-$R$ retrieved pages.

\begin{table*}[t]
\centering
   \caption{R-precision / DR$_{OOD}$ for query variations on the FEV and NQ dev data. Significant performance degradation with respect to the corresponding  I.I.D. setting is denoted as `$-$' ($\textit{p-value} \leq 0.05$).}
   \renewcommand{\arraystretch}{0.85}
   \setlength\tabcolsep{2.3pt}
  	\begin{tabular}{l  c c c c c | c c c c c }
  \toprule
  \multicolumn{1}{c}{\textbf{Dataset}} & \multicolumn{5}{c|}{\textbf{FEV}} & \multicolumn{5}{c}{\textbf{NQ}} \\
      \textbf{Model} & \textbf{Original} & \textbf{Misspelling} & \textbf{Naturality} & \textbf{Order} & \textbf{Paraphrasing} & \textbf{Original} & \textbf{Misspelling} & \textbf{Naturality} & \textbf{Order} & \textbf{Paraphrasing} \\ 
       \midrule
DPR  &  52.9 & 24.1/-54.4 & 32.4/-38.8 & 22.3/-57.8 & 34.8/-34.2 & 53.9 & 25.6/-52.5 & 31.8/-41.0 & 31.0/-42.5 & 44.6/-17.3 \\
BART  & 79.6 & 20.7/-74.0 & 38.3/-51.9 & 22.1/-72.2 & 34.7/-56.4 & 48.9 & 26.2/-46.4 & 39.1/-20.0 & 32.8/-32.9 & 43.4/-11.2 \\
C.B.  & \textbf{81.4} & \textbf{26.0}/-68.0 & \textbf{41.8}/--48.6 & \textbf{27.7}/-66.0 & \textbf{40.6}/-50.1 & \textbf{57.6} & \textbf{28.1}/-51.2 & \textbf{39.2}/-31.9 &  \textbf{36.1}/-37.3 & \textbf{50.1}/-13.0 \\
\bottomrule
    \end{tabular}
    \vspace*{-2mm}
   \label{table: query ood}
\end{table*}

\begin{table}[t]
\vspace*{-1mm}
\centering
   \caption{R-precision / DR$_{OOD}$ for unforeseen query types on the open domain QA dataset of KILT. The trn and tst refer to train and test respectively.}
   \renewcommand{\arraystretch}{0.85}
   \setlength\tabcolsep{8.5pt}
  	\begin{tabular}{l l c c c c c }
  \toprule
   \textbf{Model} & trn \textbackslash{} tst & \textbf{NQ} & \textbf{HoPo} & \textbf{TQA} \\ 
       \midrule
\multirow{3}{*}{DPR} 
& \textbf{NQ} & 53.9 & 23.1/-11.5 & 29.2/-31.8 \\ 
& \textbf{HoPo} & 41.2/-23.6 & 26.1 & 26.3/-38.6 \\
& \textbf{TQA} & 42.3/-21.5 & 21.8/-16.5 & 42.8 \\ \hline
\multirow{3}{*}{BART} 
& \textbf{NQ} & 48.9 & 36.4/-12.5 & 50.7/-21.3 \\ 
& \textbf{HoPo} & 18.8/-61.6 & 41.6 & 46.8/-27.3 \\
& \textbf{TQA} & 26.5/-45.8 & 35.2/-15.4 & 64.4 \\ \hline
\multirow{3}{*}{C.B.} 
& \textbf{NQ} & \textbf{57.6} & 47.0/-7.3 & 52.7/-25.7\\ 
& \textbf{HoPo} & 33.4/-42.0 & \textbf{50.7} & 48.6/-31.5 \\
& \textbf{TQA} & 32.9/-42.9 & 44.7/-11.8 & \textbf{70.9} \\
\bottomrule
    \end{tabular}
    \vspace*{-2mm}
   \label{table: cross dataset}
\end{table}

\begin{table}[t]
\centering
\vspace*{-2mm}
   \caption{R-precision / DR$_{OOD}$ for unforeseen tasks on the 5 KILT task-mixed datasets. The trn and tst refer to train and test respectively.}
   \renewcommand{\arraystretch}{0.85}
   \setlength\tabcolsep{1.3pt}
  	\begin{tabular}{l l c c c c c c }
  \toprule
   \textbf{Model} &  trn \textbackslash{} tst & \textbf{FC} & \textbf{EL} & \textbf{SF} & \textbf{QA} & \textbf{Dial.} \\ 
       \midrule
\multirow{5}{*}{DPR} 
& \textbf{FC} & 52.9 & 0.8/-61.9 & 15.2/-21.1 & 22.3/-37.9 & 24.0/-5.9 \\ 
& \textbf{EL} & 46.8/-11.5 & 2.1 & 13.6/-29.5 & 15.9/-55.7 & 13.6/-46.7 \\
& \textbf{SF} & 45.4/-14.2 & 0.2/-90.5 & 19.3 & 20.2/-43.7 & 10.0/-60.8 \\
& \textbf{QA} & 50.8/-4.0 & 0.6/-71.4 & 12.1/-37.3 & 35.9 & 22.5/-11.8 \\
& \textbf{Dial.} & 48.2/-8.9 & 0.7/-66.7 & 10.3/-46.6 & 25.1/-30.1 & 25.5 \\ \hline
\multirow{5}{*}{BART} 
& \textbf{FC} & 79.6 & 10.0/-88.9 & 81.1/-3.3 & 41.5/-29.7 & 48.1/-3.2 \\ 
& \textbf{EL} & 68.6/-13.8 & 90.1 & 77.4/-7.7 & 24.6/-58.3 & 21.5/-56.7 \\
& \textbf{SF} & 65.6/-17.6 & 3.7/-95.9 & 83.9 & 36.3/-38.5 & 17.1/-65.6 \\
& \textbf{QA} & 76.4/-4.0 & 9.8/-89.1 & 77.6/-7.5 & 59.0 & 45.0/-9.5 \\
& \textbf{Dial.} & 71.8/-9.8 & 8.6/-90.5 & 69.2/-17.5 & 40.2/-31.8 & 49.7 \\ \hline
\multirow{5}{*}{C.B.} & \textbf{FC} 
& \textbf{81.4} & 9.3/-89.7 & 82.2/-2.5 & 48.9/-19.4 & 46.8/-19.3 \\ 
& \textbf{EL} & 68.6/-15.7 & \textbf{90.7} & 62.7/-25.6 & 38.4/-36.7 & 33.0/-43.1 \\
& \textbf{SF} & 68.1/-16.3 & 4.9/-94.6 & \textbf{84.3} & 43.7/-28.0 & 25.3/-56.4 \\
& \textbf{QA} & 79.2/-2.7 & 8.1/-91.1 & 78.7/-6.6 & \textbf{60.7}& 46.3/-20.1 \\
& \textbf{Dial.} & 74.1/-9.0 & 6.9/-92.4 & 77.7/-7.8 & 49.3/-18.8 & \textbf{58.0} \\
\bottomrule
    \end{tabular}
   \label{table: cross task}
\end{table}

\vspace{-2mm}
\section{Result}
We examine the empirical results in the I.I.D. setting and the three OOD settings sequentially: query variations, unforeseen query types, and unforeseen tasks.

\vspace{-2mm}
\subsection{The Overall I.I.D. Result}
We compare the retrieval models on the KILT benchmark.
From Table \ref{table:Baseline}, we can observe that the generative retrieval models significantly outperform dense retrieval models like DPR across all the datasets, indicating that combining the retrieval components into a unified model benefits effective corpus indexing.
CorpusBrain consistently outperforms BART on all five tasks, demonstrating that the adequately well-designed pre-training tasks for generative retrieval contribute to improving document understanding for generative retrieval models.

\vspace{-1mm}
\subsection{Analysis of OOD Generalizability on Query Variations}
Firstly, from Table \ref{table: query ood}, we provide a comprehensive performance analysis of all the retrieval models.
We observe a significant effectiveness drop for query variations in all retrieval models. 
The results indicate that both dense retrieval models, as well as generative retrieval models, are not robust to query variations, which complements the findings from previous work \cite{penha2022evaluating}.

When we compare the generalizability of the dense and generative retrieval models, we find that the generative retrieval models perform particularly poorly on \textbf{Misspelling} and \textbf{Order}.
One possible explanation would be that the generative retrieval models generate document identifiers autoregressively based on the query, so query quality and word order greatly impact the generation effect.
When we look at the performance of generative retrieval models, we can find that the CorpusBrain has better R-precision than BART, indicating that pre-training tasks tailored for generative retrieval help the model adapt better to query variations.

\vspace{-2mm}
\subsection{Analysis of OOD Generalizability on Unforeseen Query Types}
When we look at the overall performance of all the retrieval models from Table \ref{table: cross dataset}, we can observe that as the shift of query types distributions, the performance of all models decreases significantly.
This suggests that, even under the same task, retrieval models face challenges of poor OOD generalizability.
Consequently, it is important to consider the OOD performance for these unseen query types.

Comparing the performance of dense and generative retrieval models, we find that generative retrieval models exhibit worse generalizability on web search queries in the NQ dataset.
It indicates that, in terms of generalizability performance on unforeseen query types, generative retrieval models behave differently and merit separate studies.
Furthermore, we observe that CorpusBrain demonstrates better generalizability than BART on unforeseen query types.
This could be attributed to the pre-training process of CorpusBrain, which effectively encodes relevant information for a given corpus to cope with potentially unknown queries, thereby enhancing its stability when encountering unforeseen query types.

\vspace{-2mm}
\subsection{Analysis of OOD Generalizability on Unforeseen Tasks}
Examining the overall performance of all retrieval models in Table \ref{table: cross task}, we observe that the generalizability defects for unforeseen tasks are common among models.
In the entity linking (EL) task, the models' generalization performance drops significantly, likely due to the task's distinct format compared to the others.

When we observe the performance between dense and generative retrieval models, we find that, in general, generative retrieval models have higher $DR_{OOD}$ on slot-filling (SF) task.
This could be because the format of this downstream task aligns with the pre-training tasks of the backbone generative models.
Comparing BART and CorpusBrain from the generative retrieval models, we observe that CorpusBrain outperforms BART in most (13 out of 20) unforeseen task scenarios.
This may be attributed to the pre-training tasks of CorpusBrain.
CorpusBrain includes three tasks: Inner Sentence Selection (ISS), Lead Paragraph Selection (LPS), and Hyperlink Identifier Prediction (HIP). 
ISS models the semantic granularity differences between queries and documents in various retrieval requirements, helping to bridge the gap between different downstream tasks. 
This finding is consistent with the original analysis of CorpusBrain \cite{chen2022corpusbrain}.

\vspace{-2mm}
\section{Conclusion}

In this paper, we analyzed the out-of-distribution robustness of several representative generative and dense retrieval models on the KILT benchmark. 
Specifically, we proposed three perspectives to define the out-of-distribution robustness. 
The results showed that generative retrieval models expose significant vulnerabilities in OOD robustness.
More research efforts are needed to develop robust generative retrieval models. 

Our work highlights the need to create benchmarks that include various OOD perspectives to better understand the generative retrieval models' robustness.
In future work, we will introduce more generative retrieval models with more datasets to further explore the OOD robustness.
we would also try to apply the findings to improve the robustness of existing generative retrieval models.

\clearpage
\bibliographystyle{ACM-Reference-Format}
\balance
\bibliography{references}

\end{document}